\documentclass[aps,prl,twocolumn]{revtex4}
\usepackage{amsfonts}
\usepackage{amsmath}
\usepackage{amssymb}
\usepackage{graphicx}%
\usepackage{color}
\usepackage{hyperref}
\usepackage{calc}
\begin{document}

\title{Reply to the Comment on \\``General Non-Markovian Dynamics of Open Quantum System"}
\author{Wei-Min Zhang$^1$, Ping-Yuan Lo$^1$, Heng-Na Xiong$^1$, Matisse Wei-Yuan Tu$^1$ and Franco Nori$^{2,3}$}
\affiliation{$^1$Department of Physics, National Cheng Kung University, Tainan, 70101 Taiwan \\
$^2$Center for Emergent Matter Science, RIKEN, Saitama 351-0198, Japan \\
$^3$Physics Department, The University of Michigan, Ann Arbor, Michigan, 48109-1040, USA}

\date{July 28, 2014}

\maketitle

The letter \cite{Zhang12} presents three examples.
For the steady-state solution of the first example, i.e., the dissipationless 
part of Eq.(12) in \cite{Zhang12}, the 2nd version of the Comment \cite{McC13} claimed that 
``this [dissipationless] regime exists if and only if the total Hamiltonian is 
unbounded from below, casting serious doubts on the usefulness of this result." 
In the following, we shall show that this Comment is again incorrect.

The total Hamiltonian used in the first example in \cite{Zhang12} is 
$H_{\rm tot} \!= \omega_s a^\dag a + \sum_k \omega_k b^\dag_k b_k
+ \sum_k V_k (a^\dag b_k + b^\dag_k a)$.  Diagonalizing $H_{\rm tot}$ leads to 
$H_{\rm tot} \!= \omega_b C^\dag C + \sum_k \omega'_k D^\dag_k D_k $,  where  
$\omega_b  \!=\omega_s \!- \!\int_0^\infty \!\! d\omega \frac{J(\omega)}{\omega\!-\!\omega_b}$ 
is the renormalized mode of the system, and $\omega_{k}\! <\!\omega'_k \!< \!\omega_{k+1}$.
The operators $\{C^\dag, D^\dag_k\}$ are all normal 
modes of the total system after a Bogoliubov transformation from the  
basis $\{a^\dag, b^\dag_k\}$ \cite{SM}, and $C^\dag$ is just
the single-excitation given in \cite{McC13}.  In the continuous limit, $\omega'_k=\omega_k$. 
Thus, the first mistake made in \cite{McC13} is that at operator level,
$H_{\rm tot}$ is not given only by $C^\dag$.

Secondly, Arai and Hirokawa proved \cite{AH00} that the spectrum of the above
Hamiltonian in the strong-coupling regime is unbound from below when the particle 
number in the renormalized mode $\omega_b$ is unbound.  However,  
due to the conservation of the total particle number, $[H_{\rm tot}, 
N_{\rm tot}] \!=0$, where $N_{\rm tot}\! = a^\dag a + \sum_k b^\dag_k b_k$, the total 
Hamiltonian can be written as a direct sum of decomposed Hamiltonians. Each 
decomposed Hamiltonian with fixed total particle number always has a lower bound 
for arbitrary coupling $V_k$ \cite{Hir14}.  A similar ``unbound" ground state energy
also exists for the total Hamiltonian of a Dirac particle in QED, where 
the possible trouble from the unbound ground-state energy in QED is avoided due to the total 
momentum conservation, i.e., the decomposed Hamiltonian with fixed total momenta has 
a lower bound for arbitrary QED coupling \cite{Sa05}. Thus, the second mistake made in \cite{McC13} 
was not to consider the important role of the particle number conservation.

Furthermore, because $H_{\rm tot}$ is not simply given by the single-excitation $C^\dag$,  
the third mistake made in \cite{McC13} is that the possible energy divergence 
claimed in \cite{McC13} is not applicable to the non-Markovian 
dynamics studied in \cite{Zhang12}. 
Non-Markovian dynamics relies on the initial states of the total system.  
The exact master equation  formalism given in \cite{Zhang12} requires that  the initial 
states of the total system {\it must be a direct product state} between the system 
and its environment. These states always carry a positive-definite total energy.  
This decoupling condition {\it must} be obeyed for any 
exact master equation derived from the Feynman-Vernon influence 
functional \cite{Fey63}. Otherwise one cannot carry out the influence 
functional and thereby would be unable to derive the exact master equation.  Thus, due to the total 
energy conservation, the total Hamiltonian in our study  
\cite{Zhang12} is always positive-definite.
 
To be more specific, let us begin with the valid initial state: $|\psi(t_0)\rangle \!\!= \!
a^\dag |0,\{0_k\}\rangle $, in which the system initially contains one particle, and the 
environment is in its vacuum, i.e., the system and the environment are
initially decoupled \cite{Fey63}. The corresponding energy of the total system is just the
energy carried by the particle in the initial state, i.e., $ E_{\rm tot}\!=\!\omega_s \!>\!0$. 
Solving exactly the Schr\"{o}dinger equation with this initial state, 
the steady state of the total system is $|\psi(t \rightarrow \infty)\rangle = 
\big[e^{-i\omega_b t}{\cal Z}\big(a^\dag + \sum_k \frac{V_k}
{\omega_b-\omega_k}b^\dag_k  \big) 
+ \sum_k e^{-i\omega_kt} [\omega_k-\omega_s - \Delta(k) +i\gamma(k)]^{-1} 
b^\dag_k \big]|0,\{0_k\}\rangle $ which is a {\it superposition} of the renormalized mode 
 $\omega_b$ of the system {\it plus} all other possible modes $\omega_k$ of the environment, 
where the first term gives the dissipationless part of the system in \cite{Zhang12}. 
The derivation of this result is given in \cite{SM}. 
In the strong-coupling regime, $\omega_b$ is negative, as shown in \cite{Zhang12}, 
but the energy of the total Hamiltonian is positive, $E_{\rm tot}\!=\!\omega_s\! >\!0$,
because the total Hamiltonian and the total particle number are conserved during 
the time evolution. 
Let us now extend the above solution to the initial states $|\psi_n(t_0)\rangle
\propto (a^\dag)^n |0,\{0_k\}\rangle $, where $n=1,2,3, \cdots$, can be any arbitrary integer.
The corresponding steady state of the total system is $|\psi_n(t \rightarrow \infty)\rangle 
\propto \big[e^{-i\omega_b t}{\cal Z} \big(a^\dag + \sum_k \frac{V_k}
{\omega_b-\omega_k}b^\dag_k  \big)    + \sum_k e^{-i\omega_kt} [\omega_k-
\omega_s - \Delta(k) +i\gamma(k) ]^{-1} b^\dag_k \big]^n|0,\{0_k\}\rangle $. The total 
energies of all these states $E_{\rm tot,n}\!=\!n\omega_s\! >\!0$ ({\it positive-definite}).  
In fact, for all valid initial states and the corresponding exact solutions of the 
master equation given in \cite{Zhang12}, the total Hamiltonian is always {\it positive-definite}.  
This is consistent with the fact that the total Hamiltonian with fixed total particle 
numbers has a lower bound, due to the total particle number conservation. 
Thus, the above-quoted criticism in \cite{McC13} is obviously incorrect.  
The Comment \cite{McC13} ignored the validity of the exact master equation 
derived from the Feynman-Vernon influence functional to reach  an incorrect conclusion.

The only correct part in the Comment is the last part of \cite{McC13}, 
where  they pointed out that the problem exists in {\it their} own works, i.e., Refs.~3 
and 4 in \cite{McC13} for the quantum Brownian motion (QBM). This is because the
QBM used a system-environment coupling $H_{\rm int}=\sum_k c_kx q_k
=\sum_k c'_k(a^\dag b_k + a b^\dag_k + a^\dag b_k^\dag + a b_k)$, 
which breaks the conservation of the  total particle number, $[H_{\rm int}, N_{\rm tot}]\neq 0$.  
Because of this, the QBM will cause 
both dynamical and thermodynamic instabilities in the strong-coupling regime, 
as claimed in their Comment \cite{McC13}. The authors of \cite{McC13} 
did not realize 
that the instabilities of the QBM in the strong-coupling regime come from the breakdown of  
the total particle number conservation. Our studies focus on systems preserving the 
total particle number conservation and therefore do not have such problem (see 
more discussions in \cite{Zhang14}).  They mistakenly believe that the instability 
aspects in their own works could also be applicable to other Hamiltonians.

\end{document}